\documentclass[lettersize,journal]{IEEEtran}
\usepackage{setspace}

\usepackage{amsmath,amsfonts}
\usepackage{color}
\usepackage{soul}
\usepackage{amssymb}
\usepackage{algorithm}
\usepackage{algpseudocode}
\usepackage{array}
\usepackage{textcomp}
\usepackage{stfloats}
\usepackage{subcaption}
\usepackage{url}
\usepackage{verbatim}
\usepackage{graphicx}
\usepackage{tikz}
\usetikzlibrary{arrows.meta,positioning,fit,backgrounds}
\usepackage{cite}
\usepackage{xcolor}
\usepackage{bbding}
\usepackage{hyperref}

\begin{document}


\title{Label-Free Concept Drift Assessment for Reliable AI in Emerging Wireless Applications}

\author{\IEEEauthorblockN{
    Athanasios Tziouvaras \IEEEauthorrefmark{1}, Carolina Fortuna\IEEEauthorrefmark{2} , George Floros \IEEEauthorrefmark{5}, Kostas Kolomvatsos \IEEEauthorrefmark{4}, Panagiotis Sarigiannidis\IEEEauthorrefmark{3}, Marko Grobelnik\IEEEauthorrefmark{2} and Bla\v{z} Bertalani\v{c} \IEEEauthorrefmark{2}\\
}\IEEEauthorblockA{
    \IEEEauthorrefmark{1} Business and IoT Integrated Solutions LTD, Nicosia, Cyprus, 
    \IEEEauthorrefmark{2} Jo\v{z}ef Stefan Institute, Slovenia\\
    \IEEEauthorrefmark{5} Trinity College Dublin, Dublin, Ireland, 
    \IEEEauthorrefmark{4} University of Thessaly, Lamia, Greece\\
     \IEEEauthorrefmark{3} University of Western Macedonia, Kozani, Greece\\
}

blaz.bertalanic@ijs.si
\\
}

\maketitle

\begin{abstract}

Machine learning models deployed in non‐stationary environments degrade silently, since as the input distribution drifts their accuracy decays without an error signal and without labels to reveal it. Sustaining reliable AI therefore requires a concept‐drift detector that acts as an external \emph{observer} of the deployed model, monitoring it using unlabeled operational data alone, so that an MLOps \emph{actuator} triggers retraining and redeployment only when it is warranted. This paper contributes two concept drift detectors, namely Confidence‐Filtered Pseudo‐Label Transfer (CFPT) and TabAutoDrift, which combine representation learning with statistical testing to compute an expected utility score that signals whether a deployed model should be retrained, without requiring ground‐truth labels after deployment. The detectors are evaluated on two emerging, label‐scarce wireless application domains in which post deployment ground truth is effectively unavailable, namely outdoor fingerprinting‐based localization and link‐anomaly detection. They outperform the classical detectors ADWIN, DDM, and CUSUM, attaining a drift‐detection F1‐score between 0.88 and 0.94 in the fingerprinting use case and between 0.80 and 1.00 in the link‐anomaly use case, up to 0.24 higher than the strongest classical detector. Interpreted as reliability decisions, this precision indicates that the proposed detectors signal retraining more dependably.


\end{abstract}


\section{Introduction}

Machine‐learning models are increasingly deployed as long‐lived components of automated systems, where they operate on data streams whose statistical properties evolve over time as depicted at the top of Figure~\ref{fig:overview}. This non‐stationarity, commonly termed concept drift, constitutes a fundamental obstacle to reliable AI, since a model trained on historical data gradually loses accuracy as the environment changes, with no intrinsic error signal to announce the degradation~\cite{1}. Maintaining this reliability calls for a feedback control loop around the model, in the spirit of self-adaptive ML systems~\cite{gheibi2021applying} and MLOps practice~\cite{kreuzberger2023mlops}, following an observe $\rightarrow$ decide $\rightarrow$ act pattern. As depicted in Figure~\ref{fig:overview}, the deployed model is the \emph{System Under Observation} (SUO); a concept-drift detector is the \emph{Observer} (O) that monitors the SUO and decides when its learned assumptions no longer hold; and an MLOps pipeline is the \emph{Actuator} (A) that executes the decision by retraining and redeploying the model. The central difficulty is that this detection must be performed without labels, because in most realistic deployments the ground‐truth outcomes for post‐deployment data are delayed, sparse, or entirely unavailable~\cite{liu2023leaf}, so that the very signal on which a supervised drift test would rely is absent. Figure~\ref{fig:overview} summarizes this setting together with the structure shared by the two proposed detectors realizing the Drift observer.

\begin{figure}[t]
\centering
\begin{tikzpicture}[
  font=\scriptsize,
  >={Stealth[length=1.5mm]},
  box/.style={draw, rounded corners, align=center, inner sep=3pt},
  data/.style={box, fill=black!5, text width=50mm},
  model/.style={box, fill=orange!18, text width=50mm},
  blk/.style={box, fill=blue!8, text width=50mm},
  ops/.style={box, fill=green!14, text width=50mm},
  arr/.style={->, semithick},
]
\node[data] (stream) {Non-stationary data stream $D_1$}; 
\node[model, below=4mm of stream] (m0) {\textbf{System Under Observation}: model $M_0$ deployed in production\\ \textit{accuracy decays silently, no error signal}};
\node[blk,dashed, below=9mm of m0] (train) {\textbf{Drift observer offline initialization:} \\ Train the drift observer on $D_0,L_0$;\\ Collect macro-F1 $F$};
\node[blk, dashed, below=4mm of train] (retrain) {\textbf{Drift observer online ingestion:} \\ Re-train the drift observer on new data $D_1$;\\ Collect macro-F1 $F_t$};
\node[blk, dashed, below=4mm of retrain] (util) {\textbf{Drift observer significance testing:}\\ Calculate utility $U=\tfrac{1}{n}\sum_i |F_i-F_t^{i}|$;\\ Trigger $s$ if $U \ge \theta$};
\node[ops, below=7mm of util] (mlops) {\textbf{Actuator}:\\ MLOps pipeline acting on \textbf{Observer} decision.};
\begin{scope}[on background layer]
  \node[draw, rounded corners, fill=blue!3, fit=(train)(retrain)(util), inner sep=3mm] (det) {};
\end{scope}
\node[anchor=south west, font=\scriptsize\itshape] at (det.north west) {\textbf{Drift observer} (label-free)};
\draw[arr] (stream) -- (m0);
\draw[arr] (m0) -- (train);
\draw[arr] (train) -- (retrain) node[pos=0.4,right]{transfer};
\draw[arr] (retrain) -- (util);
\draw[arr] (util) -- (mlops) node[pos=0.7,right=1pt]{signal $s$};
\draw[arr] (mlops.west) -- ++(-7mm,0) |- (m0.west) node[pos=0.28,left]{redeploy};
\end{tikzpicture}
\caption{Overview of a set-up and the structure of the proposed detectors. 
}
\label{fig:overview}
\end{figure}

This challenge is particularly acute in AI‐native wireless networks, which embed AI/ML models from the radio access network (RAN) to the core~\cite{wilhelmi10726906, saimler10597102} in order to deliver automation and dynamic adaptation. As the network environment evolves through changing user behaviours, new application deployments, varying traffic, and subtle shifts in the radio frequency landscape, the operational data diverge from the training data~\cite{liu10670198}, and models that operate under outdated assumptions degrade the quality of service and threaten network stability~\cite{yungaicela2024misconfiguration}. Wireless environments thus constitute a demanding and high‐stakes instance of the general reliability problem, as their data distributions are inherently non‐stationary and true annotations are extremely scarce after deployment.

We investigate this problem in two emerging, label‐scarce wireless domains in which it is especially pronounced, namely localization fingerprinting~\cite{14} and wireless anomaly detection~\cite{9715175}. In fingerprinting‐based localization, models map historical radio measurements to physical locations, yet infrastructure changes, device heterogeneity, and evolving user patterns shift these mappings over time, while obtaining true positions for live measurements requires physical ground‐truthing. In wireless anomaly detection, models trained on normal operational data fail to identify new fault types once their assumptions become outdated, and fault annotations are rare and costly to obtain. Both domains therefore exhibit precisely the regime in which a drift detector cannot rely on supervised signals.

In this work, we contribute a model‐agnostic, label‐free concept‐drift detector as Observer that monitors a deployed model as SUO, using only unlabeled operational data and signals when it should be retrained, thereby providing the trigger that an MLOps pipeline consumes in order to retrain and redeploy the model as depicted in Figure~\ref{fig:overview}. The main contributions of this paper are as follows.

\begin{itemize}
    \item Two model‐agnostic, label‐free drift‐assessment detectors, namely Confidence‐Filtered Pseudo‐Label Transfer (\textit{CFPT}) and \textit{TabAutoDrift}, which combine representation learning and statistical testing, operate on both tabular and time‐series data, and require no ground‐truth labels after deployment.
    \item A bounded expected‐utility score that converts drift assessment into a retraining signal, indicating when a deployed model has degraded sufficiently to warrant retraining by an MLOps pipeline. We show that this signal is both accurate, in terms of a high detection F1‐score, and precise as a reliability trigger, in terms of few false and no missed retraining alarms.
    \item A validation on two emerging, label‐scarce wireless domains using real‐world data, in which the detectors outperform the classical detectors ADWIN, DDM, and CUSUM, attaining a drift‐detection F1‐score between $0.88$ and $0.94$ for the \textit{fingerprinting} use case and between $0.8$ and $1.00$ for the \textit{links} use case, while requiring fewer and more transportable hyperparameters. Raising the retraining alarm lies within the scope of this work, whereas performing the retraining is left to the operator.
\end{itemize}
Table \ref{tab:nomenclature} encapsulates the core nomenclature we use to distinguish the different datasets and their corresponding labels.

\begin{table}[h]
    \centering
\caption{ Core nomenclature used throughout the paper.}
\label{tab:nomenclature}
\resizebox{.49\textwidth}{!}{
\begin{tabular}{c|c}
\hline
$X_k,\ y_k$ & Feature vector and its label.        \\
$D_0, L_0$  & Original training set and its labels. \\
$M_0$       & A model that is trained on $D_0, L_0$. \\
$D_1$       & A new unlabeled dataset which is assessed for drift.  \\
$L_1$       & Pseudo-labels assigned to $D_1$, using the proposed detectors.\\
$L_1^{gt}$  & Manually annotated ground-truth labels for $D_1$, \\
& used only for offline evaluation.\\
\hline
\end{tabular}
}
\end{table}

 This paper is organized as follows. Section \ref{sec:related} summarizes related work, Section \ref{sec:drift} provides the background and problem formulation, and Section \ref{sec:concept_drift_method} introduces the proposed detectors. Section~\ref{sec:methodology} details the evaluation methodology, Section \ref{sec:results} presents the results, while Section \ref{sec:conclusion} concludes the paper.

\section{Related work}
\label{sec:related}
Although ``data drift'' and ``concept drift'' are often used interchangeably, we adopt the taxonomy of \cite{moreno2012unifying} (covariate, prior-probability, concept, and dataset shift; detailed in Section~\ref{sec:drift}). Our detectors are agnostic to the drift type, flagging any distribution change large enough to degrade the deployed model and warrant its retraining; in our wireless use cases such change stems primarily from the emergence of new classes.

\subsection{Concept Drift Detection Techniques}
\label{sec:drift_tech}

Concept drift denotes changes in the input--output relationship over time that degrade models if left unaddressed~\cite{1}. Existing detectors fall into three broad categories: statistical change detection, window-based distribution methods, and deep-learning-based methods.

\subsubsection{Statistical change detection methods} These apply hypothesis tests or control charts. The Drift Detection Method (DDM)~\cite{ddm-cmp} monitors the input data stream and raises an alarm when the amount drifted samples exceeds an expected limit. Sequential tests such as Page-Hinkley~\cite{page-cmp} and CUSUM~\cite{page-cmp} detect mean shifts in performance metrics, treating drift detection as control-process change detection. Such detectors are lightweight and flag abrupt shifts quickly but struggle with gradual drifts.

\subsubsection{Window-based distribution methods} These compare data distributions across time windows. Adaptive Windowing (ADWIN)~\cite{adwin-cmp} maintains a sliding window that grows or shrinks to keep old and new data statistically consistent, and later detectors such as STEPD~\cite{stepd-cmp} use reference and test windows. Their challenge is window sizing: too small windows over-react to noise, too large a windows delay detection.

\subsubsection{Deep learning based methods} These mostly use autoencoders~\cite{jaworski2020concept}: An autoencoder trained on an initial window flags drifts when its reconstruction error on new data rises sharply, capturing sudden drifts. An alternative integrates the detector into the model itself~\cite{yuan2022recent}, but this must be designed in and trained jointly with the model, so it is not model-agnostic.

Our two methods address these shortcomings: both are model-agnostic, detect abrupt and gradual drift effectively, and rely on fewer, more transportable hyperparameters than the baselines, which require window sizes, forgetting factors, and sensitivity thresholds recalibrated per deployment.

\subsection{Unsupervised and Label-Free Drift Detection}
\label{sec:rw_unsup}
Because labels are frequently unavailable after deployment, a growing body of work detects drift without them by testing whether the input distribution has changed. Classifier-based two-sample tests train a discriminator to separate reference from recent data and infer drift from its accuracy, as in the Discriminative Drift Detector (D3)~\cite{gozuacik2019d3} and the dataset-shift study of Rabanser et al.~\cite{rabanser2019failing}. Distribution-distance approaches monitor divergences such as the incremental Kolmogorov--Smirnov test~\cite{dosreis2016fast} or maximum-mean-discrepancy tests, while label-shift methods estimate class-prior changes without target labels~\cite{lipton2018detecting}. Our detectors, which realize the Observer of Figure~\ref{fig:overview}, belong to this label-free family but differ by coupling the change test to the retraining decision through a bounded expected-utility score, rather than reporting a distance in isolation, and by operating uniformly on tabular and time-series wireless data.

\subsection{Concept Drift Detection in Wireless}
\label{sec:drift_tech_wireless}

In wireless settings, most work targets cybersecurity and intrusion detection~\cite{SHYAA2024109143, suarez2023survey}, stressing drift detection in heterogeneous networks for reliable intrusion detection~\cite{andresini2021insomnia,nitish2024class}, using multiple statistical tests to localize drift from changes in packet rates and protocols~\cite{chu2024intrusion}, or attributing drift to user-behaviour changes such as an intruder takeover~\cite{soltani2024multi}. These provide a solid foundation but are tightly coupled to intrusion detection and use-case specific, which hinders reuse, and many require manual recalibration that is costly or infeasible at scale.

By contrast, little work targets outdoor fingerprinting localization or link-level anomaly monitoring, two domains where data evolve unpredictably and true locations or fault annotations are extremely sparse, the label-scarce regime that most stresses a label-free detector. We address this gap with two task-agnostic detectors that operate uniformly on both domains, need substantially less calibration than standard detectors, and use hyperparameters that require no dataset-specific tuning.

\section{Background and Problem Formulation}
\label{sec:drift}
\label{sec:statement}

The terminology of ``data'' and ``concept'' drift varies across works~\cite{1,2,3}; we adopt the taxonomy of \cite{moreno2012unifying}. Let a model be trained on a dataset $D_i=\{d_0,\dots,d_j\}$, with $d_j=\{X_k,y_k\}$, drawn from a distribution $F_{0,j}(X,y)$. Drift occurs when subsequent data $d_{j+1},\dots$ follow a different distribution, that is, when there exists a $j$ such that $P_j(X,y)\neq P_{j+1}(X,y)$. Since $P_j(X,y)=P_j(X)\,P_j(y|X)$, this can be written as
\begin{equation}
\label{eq:distr_uneq}
    \exists\, j:\; P_j(X)\,P_j(y|X) \neq P_{j+1}(X)\,P_{j+1}(y|X).
\end{equation}
Following \cite{moreno2012unifying}, four cases are distinguished: \emph{covariate} (or feature) shift, where $P(X)$ changes while $P(y|X)$ is fixed; \emph{prior-probability} shift, where $P(y)$ changes while $P(X|y)$ is fixed; \emph{concept} shift, where $P(y|X)$ changes while $P(X)$ is fixed; and \emph{dataset} shift, where both $P(X)$ and $P(y|X)$ change. Our detectors are agnostic to which case occurs and flag any change large enough to warrant retraining.

\begin{figure}[t]
    \centering
    \includegraphics[width=0.3\textwidth]{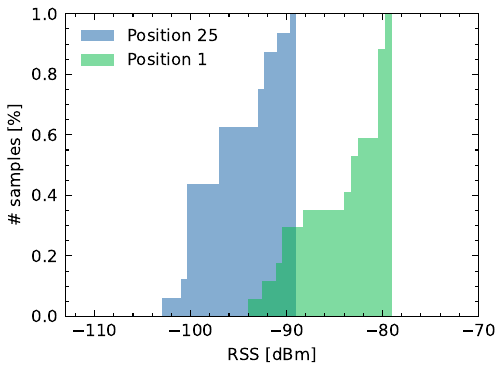}
    \caption {An example of the distributional gap underlying concept drift in the \textit{fingerprinting} dataset, shown between two different positions.}
    \label{fig:fingerprinting:presentation}
\end{figure}

\begin{figure}[t]
    \centering 

    \begin{subfigure}{0.24\textwidth}
        \centering
        \includegraphics[width=\linewidth]{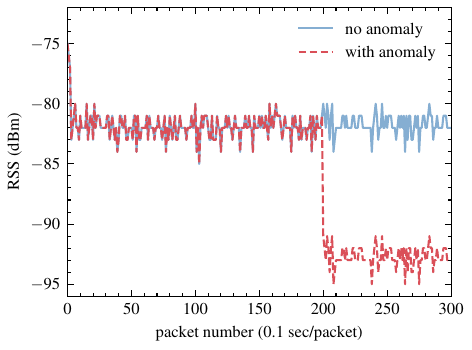}
        \caption{A first anomaly type}
        \label{fig:sub1}
    \end{subfigure}
    \begin{subfigure}{0.24\textwidth}
        \centering
        \includegraphics[width=\linewidth]{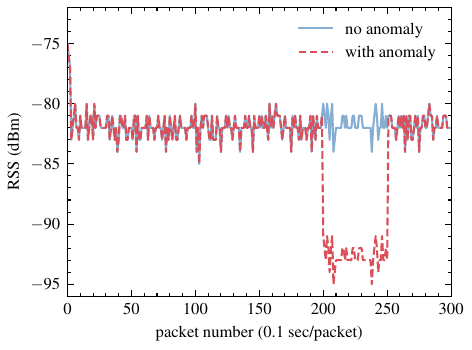}
        \caption{A second anomaly type}
        \label{fig:sub2}
    \end{subfigure}

    \vspace{0.5cm} 

    \begin{subfigure}{0.24\textwidth}
        \centering
        \includegraphics[width=\linewidth]{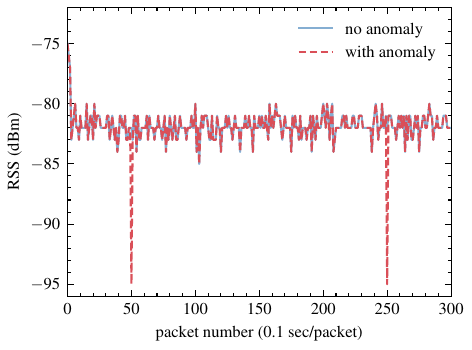}
        \caption{A third anomaly type}
        \label{fig:sub3}
    \end{subfigure}
    \begin{subfigure}{0.24\textwidth}
        \centering
        \includegraphics[width=\linewidth]{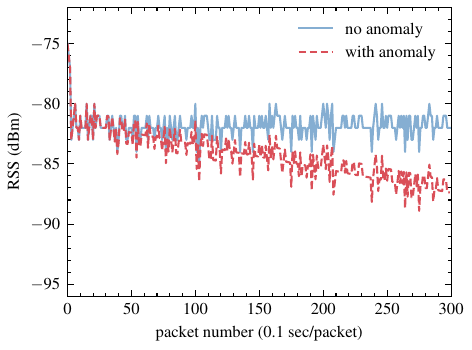}
        \caption{A fourth anomaly type}
        \label{fig:sub4}
    \end{subfigure}

    \vspace{0.5cm} 

    \begin{subfigure}{0.24\textwidth}
        \centering
        \includegraphics[width=\linewidth]{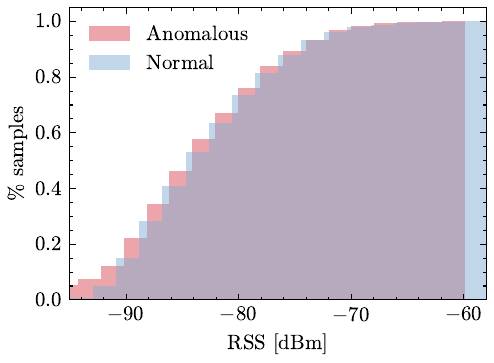}
        \caption{Data distribution between Normal and Anomalous links}
        \label{fig:sub5}
    \end{subfigure}

    \caption{An example of concept drift between normal and anomalous traces in the \textit{links} dataset.}
    \label{fig:links:presentation}
\end{figure}

We address drift detection in the operational sense: a deployed model must be retrained whenever the incoming distribution changes enough to degrade its performance. In the two wireless use cases considered here, drift is induced by the emergence of new classes, which shifts $P(X)$ and, because new and existing classes overlap in feature space, also $P(y|X)$; the induced drift is therefore a \emph{dataset shift}. Formally, the monitored model $M_0$ is fitted on the original training set $D_0$, drawn from $P_0(X,y)$, by minimizing an empirical risk. After deployment, unlabeled examples $\{(x_t)\}_{t=1}^\infty$ arrive, and at some unknown time $t^*$ the distribution shifts to $P_i\neq P_0$, so that $(x_t)\sim P_0$ for $t<t^*$ and $(x_t)\sim P_i$ for $t\ge t^*$. We cast detection as a sequential decision task and define a detector
\begin{equation*}
      \mathcal{A}\colon \{(x_i)\}_{i=1}^t \;\longmapsto\; s_t \in \{0,1\},
\end{equation*}
where $s_t=1$ triggers retraining of $M_0$, and the detector must issue $s_t$ from observed data alone, without access to future data or to labels for the incoming stream.

The objective of this work is to design $\mathcal{A}$ that a) triggers the retraining of model $M_0$ when concept drift occurs, b) controls false positives to avoid unnecessary retraining, and c) maintains high recall to prevent prolonged degradation of $M_0$. For $M_0$, we consider two different datasets that correspond to the two distinct real-world scenarios. Both leverage supervised classification tasks, one performed on a fingerprinting localization dataset from LOG-a-TEC testbed~\cite{14} and other over a time series wireless dataset from the Rutgers WinLab testbed with injected anomalies~\cite{9715175}.

Figure~\ref{fig:fingerprinting:presentation} illustrates the measurements from two different positions in the \textit{fingerprinting} dataset, where new localization locations present a concept drift. The plot shows the normalized histograms of the Received Signal Strength (RSS) for both positions and a clear separation between the two distributions is visible. When such previously unseen positions appear after deployment, this distributional gap manifests as the drift that the model must detect and adapt to.

Figure~\ref{fig:links:presentation} illustrates four representative drift patterns in the \textit{links} dataset, each of which manifests as a change in the underlying RSSI distribution and thus constitutes a form of concept drift. In all panels the solid blue trace shows the “Normal” baseline and the dashed red trace shows the anomalous RSSI measurements. Together, these examples highlight the diversity of drift phenomena in wireless links: abrupt versus gradual, persistent versus transient, single‐event versus recurring. Additionally, as it can be seen in Figure~\ref{fig:links:presentation}e, the overall distribution changes between "Normal" and "Anomalous" links is very subtle and thus hard to distinguish. This motivates the need for a unified detection framework capable of handling all of them. 

\section{Proposed concept drift detectors}
\label{sec:concept_drift_method}

\begin{algorithm}
\caption{The \textit{CFPT} algorithm.}\label{alg1:CFPT}
\begin{algorithmic}[1]
\Require Model $M_0$ (trained on $D_0, L_0$); original dataset $D_0$ with labels $L_0$; new unlabeled batch $D_1$; confidence threshold $\tau$; rounds $n$; drift threshold $\theta$
\Ensure Retraining decision $s$; Utility score $U \in [0,1]$
\State $P \gets M_0.{predict\_probability (D_1)}$ \Comment{$M_0$ frozen}
\State $(\tilde{D}_1, L_1) \gets \emptyset$ \Comment{high-confidence pseudo-labeled subset}
\ForAll{$x \in D_1$}
    \If{$\max P[x] \geq \tau$}
        \State $(\tilde{D}_1, L_1) \gets (\tilde{D}_1, L_1) \cup \{(x,\ \arg\max P[x])\}$
    \EndIf
\EndFor
\State initialize XGBoost classifier $XGB$
\For{$i \gets 1$ to $n$} \Comment{$n$ $XGB$ training rounds}
\State train $XGB$ on $(D_0, L_0)$ for one round
\State $F^{XGB}_i \gets {macroF1}\{XGB, D_0, L_0\}$
\EndFor
\State weight transfer from $XGB$ to $XGB_t$
\For{$i \gets 1$ to $n$} \Comment{$n$ $XGB_t$ training rounds}
\State train $XGB_t$ on $(\tilde{D}_1, L_1)$ for one round
\State $F^{XGB_t}_i \gets {macroF1}\{XGB_t, \tilde{D}_1, L_1\}$
\EndFor
\State $U \gets \dfrac{1}{n}\displaystyle\sum_{i=1}^{n}\left| F^{XGB}_i - F^{XGB_t}_i\right|$ 
\If{$U \geq \theta$}
    \State $s \gets 1$ \Comment{Retrain $M_0$}
\Else
    \State $s \gets 0$ \Comment{Do not retrain $M_0$}
\EndIf
\end{algorithmic}
\end{algorithm}

In this section we propose two possible instantiations of the Drift observer from Figure~\ref{fig:overview}, namely CFPT and TabAutoDrift, designed to be deployed in production to monitor an arbitrary ML model as SUO. The workflow of each detector comprises three steps as can be seen in the figure. First, each detector is initialized in an offline fashion and trains an AI component (XGBoost in \textit{CFPT} and TabNet in \textit{TabAutoDrift}) using the $D_0$ and $L_0$. Second, the Drift observers ingest the new $D_1$ data and re-train their AI components. Thus, when monitoring the incoming data stream $D_1$ in production, our detectors do not require ground truth labels, which makes them label-free at monitoring time rather than fully unsupervised. Finally, a significance testing operation takes place, where the utility score is calculated and the Drift observer signals the actuator component. The actuator is part of an MLOps pipeline and acts upon this decision by redeploying the $M_0$ model, if necessary.

\subsection{Confidence-Filtered Pseudo-Label Transfer Method for Concept Drift Detection (\textit{CFPT})}\label{sec:CFPT}
The \textit{CFPT} realization of the Observer monitors the SUO consisting of the trained model $M_0$ and a new unlabeled batch $D_1$, which may follow a different distribution from $D_0$ and thus cause $M_0$ to underperform. The proposed method that addresses this challenge is 
elaborated in Algorithm \ref{alg1:CFPT}.

\textbf{Drift observer offline initialization.} The drift observer is initialized, as in the corresponding block of Figure~\ref{fig:overview}, in an offline manner. To this end, we train an XGBoost classifier using the dataset $D_0$, along with its $L_0$ labels as per lines 8-17 of Algorithm \ref{alg1:CFPT}. XGBoost is an ensemble learning method that utilizes Gradient Boosting to integrate multiple small models together. These small models are decision trees that are trained sequentially, each correcting the residual errors of the previous trees. According to \cite{12}, XGBoost tries to minimize the following objective function:
\begin{equation}
    \text{obj}(\phi) = \sum_{i=0}^{I} l(y_i, \hat y_i) + \sum_{j=0}^{J} \Omega (f_k)
\end{equation}
Where $l$ is the loss function that calculates the difference between the ground truth $y_i$ (in our case $L_0$) and the prediction $\hat y_i$. $\Omega(f_k)$ is a regularization term which alleviates the complexity of the model and minimizes overfitting. $\Omega(f_k)$ is calculated as follows:
\begin{equation}
    \Omega (f_k) = \gamma T + \frac{1}{2}\lambda \sum_{t=1}^{T} w_t^2
\end{equation}
where $\gamma$ is a regularization parameter, $T$ is the number of leaves in the tree $k$ and $\lambda$ is a parameter that penalizes the squared sum of tree weights $w_t^2$. For the training process, we select a small number of boosting rounds ($n=5$) to reduce the time and computational requirements of our method.

\textbf{Drift observer online ingestion}. This stage corresponds to the second block of the Drift observer, as illustrated in Figure~\ref{fig:overview}. In order to process in ingested $D_1$, we first have to secure data labels for it, which are unavailable to our system. For this reason, we leverage a pseudo-labeling method as follows: We run inference on the unlabeled $D_1$ using the frozen model $M_0$; $M_0$ is used only to predict labels and is not updated. We select the highest $\tau$ predictions of the $M_0$ i.e., the predictions with the highest confidence scores. We label them according to the $M_0$ outputs and we formulate a new labeled subset $\tilde{D}_1 \subseteq D_1$ with pseudo-labels $L_1$. This semi-supervised approach, listed in lines 1-7 of Algorithm \ref{alg1:CFPT}, creates pseudo-labels $L_1$ similarly to other works in \cite{9}, \cite{10} and \cite{11}.

In the sequel, we utilize the pseudo-labeled subset $\tilde{D}_1$, along with the generated pseudo-labels $L_1$, to initiate a model retraining operation. Model retraining is essentially a transfer learning process under which the XGBoost parameters are transferred to a new $XGB_t$ model, which is then re-trained using the new data ($\tilde{D}_1$, $L_1$). For this purpose we use the trained XGBoost classifier of the previous step (Drift observer offline initialization) and we conduct a retraining operation for $5$ rounds, while collecting statistics related with the macro F1 score of the model. Macro F1 is the average F1-score and can be computed as $F =\frac{\sum_{i=0}^{m}F_i}{m}$, where $m$ is the total number of classes in labels $L_0$, and $F_i$ is the F1-score of the XGBoost model for the class $i$. The collected F1 scores (notated as $F^{XGB}$ and $F^{XGB_t}$ in the figure) showcase the performance of the XGBoost and $XGB_t$ models with respect to the minimization of the objective function $\text{obj}(\phi)_{(D_0,L_0)}$ during the training, and the $\text{obj}(\phi)_{(D_0,L_0,D_1,L_1)}$ during the re-training operations correspondingly.

The rationale that underlies the proposed \textit{CFPT} method is the observation that the $F^{XGB_t}$ contains useful information regarding the decision to retrain the $M_0$ model. More specifically, if $F^{XGB_t} \to 0$ then the $\text{obj}(\phi)_{(D_0,L_0,D_1,L_1)}$ cannot easily converge and thus, at least one of the following statements must hold true:

\begin{enumerate}
    \item The $D_1$ contains noisy samples and the $\text{obj}(\phi)_{(D_0,L_0,D_1,L_1)}$ cannot be minimized. In such cases, the $M_0$ model fails to accurately generate the $L_1$ pseudo-labels for the $D_1$ and the XGBoost re-training operation does not achieve a good F1-score. As a result, the $M_0$ should be retrained.
    \item The $D_1$ does not contain noisy samples, but the $\text{obj}(\phi)_{(D_0,L_0,D_1,L_1)}$ cannot be minimized. In this case the $M_0$ generates the $L_1$ pseudo-labels accurately, but the XGBoost re-training operation underperforms. Since the XGBoost fails to capture the properties of both the $D_0$ and the $D_1$, we can conclude that the $D_0$ and $D_1$ contain samples from different distributions. Similarly to the previous observation, the $M_0$ should be retrained.
\end{enumerate}

\textbf{Drift observer significant testing.} In this stage, we implement the functionalities of the third block of the Drift observer, as illustrated in Figure~\ref{fig:overview}. More specifically, we assess the expected utility by leveraging the observations from the previous stages and we apply the formula of Algorithm \ref{alg1:CFPT}, line $18$. 
Since $F^{XGB} \in [0,1]$ and $F^{XGB_t} \in [0,1]$, we expect $U \in [0,1]$ as well. When $U \to 1$, the $\text{obj}(\phi)_{(D_0,L_0)}$ and $\text{obj}(\phi)_{(D_0,L_0,D_1,L_1)}$ converge at different rates, or does not converge at all, and thus, the $D_0$ and $D_1$ differ significantly. In such cases, the $M_0$ should be re-trained with the $D_1$ dataset. We proceed to generate the corresponding $s$ signal to the actuator component of Figure~\ref{fig:overview}, which redeploys the $M_0$ model, if required.

\subsection{TabNet Autoencoding for Concept Drift Detection Method (\textit{TabAutoDrift})}
\label{sec:TabNet}

Unlike CFPT, which queries $M_0$ for pseudo-labels, \textit{TabAutoDrift} derives its drift signal directly from the data $D_0$ and $D_1$. The method is 
presented in Algorithm \ref{alg2:TabAutoDrift}.

\textbf{Drift observer offline initialization.} The offline initialization (in accordance to Figure~\ref{fig:overview}) of the Drift observer is performed using TabNet~\cite{13}. TabNet is a deep learning network that works with tabular data as per lines 1-8 of Algorithm \ref{alg2:TabAutoDrift}. TabNet integrates several sequential attention layers and feature selection mechanisms to learn the most useful features from a set of input data. It can be used either as a representation learning tool to tune its weights to capture the input data features, or as standard classification model. 

During the offline initialization, we select the $D_0$ dataset, along with its labels $L_0$ and we utilize a TabNet model for classification. This is a standard supervised training task, which takes $5$ epochs. At the end of it, we collect the macro F1-score $F^{Tab}$ that captures the average F1-score of the TabNet model.


\begin{algorithm}[h]
\caption{The \textit{TabAutoDrift} algorithm.}\label{alg2:TabAutoDrift}
\begin{algorithmic}[1]
\Require Model $M_0$ (trained on $D_0, L_0$); original dataset $D_0$ with labels $L_0$; new unlabeled batch $D_1$; rounds $n$; drift threshold $\theta$
\Ensure Retraining decision $s$; Utility score $U \in [0,1]$
\State initialize TabNet classifier $TabNet$
\For{$i \gets 1$ to $n$}
    \State train $TabNet$ on $(D_0, L_0)$ for one epoch
        \State $F^{Tab}_i \gets {macroF1}\{TabNet, D_0, L_0\}$
\EndFor

\State Initialize TabNet encoder $TabNet_e$
\State Pre-train $TabNet_e$ on $D_1$ \Comment{self-supervised}
\State Initialize $TabNet_t$ with a new classifier
\State Use the $TabNet_e$ as an encoder for $TabNet_t$
\State $TabNet_t[encoder] \gets TabNet_e $ \Comment{transfer $D_1$ representation}
\For{$i \gets 1$ to $n$}
    \State train $TabNet_t$ on $(D_0, L_0)$ for one epoch 
    \State $F^{Tab_t}_i \gets {macroF1}\{TabNet_t, D_0, L_0\}$
\EndFor
\State $U \gets \dfrac{1}{n}\displaystyle\sum_{i=1}^{n}\left| F^{Tab}_i - F^{Tab_t}_i\right|$ 
\If{$U \geq \theta$}
    \State $s \gets 1$ \Comment{Retrain $M_0$}
\Else
    \State $s \gets 0$ \Comment{Do not retrain $M_0$}
\EndIf
\end{algorithmic}
\end{algorithm}

\textbf{Drift observer online ingestion.} This stage is illustrated in the second block of the Drift observer in Figure~\ref{fig:overview} and it includes a self-supervised training and a re-training operation. For the self-supervised training, we make use of TabNet's representation learning attributes and we train the model's encoder. To accomplish this, we make use of TabNet's native data masking functionality that reduces the dimensionality of input data and tunes the model's weights to select the most useful features. After the self-supervised training process is completed using the $D_1$ dataset, TabNet is able to assess each feature's importance and contribution across the entire $D_1$. From now on, we will refer to this trained TabNet encoder as $TabNet_e$.

The re-training step, as per lines 9-14 of Algorithm \ref{alg2:TabAutoDrift}, consists of two operations: For the first operation we transfer the weights of the trained encoder $TabNet_e$ to the encoder of the TabNet model which was trained during the offline initialization phase. This way re formulate a new model, named $TabNet_t$. Then, we select the dataset $D_0$, along with its labels $L_0$, as inputs and we proceed with training the $TabNet_t$ model for $5$ epochs. We also apply data masking, which is native to TabNet, to select the most important $D_0$ features. During the training we collect the $F^{Tab_t}$ which is the macro F1-score of the $TabNet_t$ model. 

\textbf{Drift observer significant testing.} We calculate the expected utility according to the equation found in the line $15$ of Algorithm \ref{alg2:TabAutoDrift}. Similarly to the \textit{CFPT} method, the utility function is bounded since $U \in [0,1]$. Thus, when $U \to 1$, the $D_0$ and $D_1$ datasets differ by a large margin, and the $M_0$ should be retrained as per lines 15-20 of Algorithm \ref{alg2:TabAutoDrift}.

\section{Evaluation methodology}
\label{sec:methodology}

In this section, we present the methodology for evaluating the proposed \textit{CFPT} and \textit{TabAutoDrift} detectors. 

\subsection{Dataset description}
\label{sec:data_descr}
In order to validate our methodology we leverage two different datasets, namely the (i) \textit{fingerprinting} \cite{14} and the (ii) \textit{links} \cite{9715175}. The fingerprinting dataset contains received signal strength (RSS) measurements made with Bluetooth Low Energy (BLE) technology, measured in dBm. The dataset consists of $505,000$ data points, organised over $25$ classes that represent 2D coordinates, which are collected during the spring and during the winter, as discussed in section \ref{sec:statement}.  We organise the data into $31$ smaller datasets each containing $16,290$ samples, as follows: The first dataset $D_0$ contains samples that belong only to the first $10$ classes of the \textit{fingerprinting} dataset (out of a total $25$ classes), and is used to train a random forest classifier, as our $M_0$ model. Datasets $D_1$ - $D_{30}$ are organised as follows: Each third dataset (i.e., $D_3$, $D_6$, $D_9$, $D_{12}$ $...$ $D_{30}$ ) contains a mix of: (i) samples stemming from new classes which do not exist in the original $D_0$; and (ii) samples that belong to the same classes as in $D_0$. We opt for different mix of percentages of drifted samples in order to evaluate the detectors under examination over various conditions. In this sense, the concept drift phenomenon exists (in different variations) in every third dataset. The rest of the datasets (i.e., $D_1$, $D_2$, $D_4$, $D_5$, $D_{7}$, $D_{8}$, ..., $D_{29}$) contain samples that belong to the classes that already exist in the $D_0$ and thus, they are not drifted. This dataset formulation is opted to simulate various scenarios of concept drifts where data from new classes are collected in the field, within different time frames.

The \textit{links} dataset contains data from $8,492$ timeseries, each one of which having $302$ data points. The dataset is labeled and is organized into $5$ classes: $1$ class for time series without an anomaly, and $4$ classes that represent different anomalies found in wireless links~\cite{9715175}. Similarly to our approach with the \textit{fingerprinting} dataset, we split the data into $9$ smaller datasets each one containing $943$ timeseries, as follows: the first dataset $D_0$ is used to train a random forest classifier, as our $M_0$ model.  $D_0$ contains samples that belong to $3$ of \textit{links} classes (out of a total $5$). Then we split the rest of the datasets $D_1 - D_8$ so that $D_1$, $D_2$ and $D_3$ contain samples from the same classes as $D_0$, while $D_4$, $D_5$, $D_6$, $D_7$ and $D_8$ contain a mix of: (i) samples from new classes; and (ii) samples from the same classes as in $D_0$. In order to pre-process each dataset to be digestible by TabNet, we treat it as a matrix of numerical features. To do this, we flatten each \textit{links} data point so that each timestep sample (e.g. $t$) becomes a feature ($X_t$). This approach gives a numerical vector of $302$ features and preserves the positional index of each measurement as a distinct input, which TabNet's attention mechanism can relate across timesteps.

Table \ref{tab:dataset_creation} summarizes our process under which several datasets were created using the \textit{fingerprinting} and \textit{links} data. There, we present the dataset splits along with the corresponding percentages of drifted samples per dataset, as discussed in the above paragraphs. Throughout the next sections, we leverage the created datasets to evaluate our detectors against techniques that constitute the current state of the art.

\begin{table}[t]
\caption{The details of the drifted and non-drifted datasets which are created based on the \textit{fingerprinting} and \textit{links}.}
\label{tab:dataset_creation}
\resizebox{.49\textwidth}{!}{
\begin{centering}
  \begin{tabular}{c|cc}
& \textit{fingerprinting} & \textit{links}\\
\hline
Total samples & $505,000$ & $8,492 \times 302$\\
$M_0$ training & $D_0$ & $D_0$\\
\hline
Drifted datasets and& $D_3(20\%), D_6(10\%), D_9(45\%),$ & $D_4(20\%), D_5(25\%), $\\
percentage of& $ D_{12}(25\%), D_{15}(15\%), D_{18}(55\%),$ & $D_6(30\%), D_7(25\%),$\\
drifted samples& $ D_{21}(15\%), D_{24}(40\%), D_{27}(35\%),$ & $D_8(15\%)$\\
& $ D_{30}(55\%)$ & \\
\hline
& $D_1, D_2, D_4, D_5, $ & \\
& $D_7, D_8, D_{10}, D_{11},$ & \\
Non-drifted datasets& $D_{13}, D_{14}, D_{16}, D_{17},$ & $D_1, D_2, D_3$ \\
& $D_{19}, D_{20}, D_{22}, D_{23},$ & \\
& $D_{25}, D_{26}, D_{28}, D_{29}$ & \\
\hline
Cause of drift & Data stemming from new & New anomaly classes are\\
& classes are mixed with $D_0$ data & added to the data\\
\end{tabular}
\end{centering}
}
\end{table}

\subsection{Experimental setup}
We developed the detectors proposed in this work using the python programming language and we implemented our AI and DNN components using the PyTorch library. The experiments were conducted using a system with NVIDIA GeForce GTX 1660 Ti 6GB (GPU) and Intel i7-10750H 2.6 GHz (CPU). All of the experiments utilise the system's GPU as a main computational resource. To support reproducibility, both datasets are publicly available from their original sources~\cite{14, 9715175}, the classical baselines are used as provided by the Frouros library~\cite{CESPEDESSISNIEGA2024101733}, and all detector hyperparameters are reported in Table~\ref{tab:baseline-param}; the implementation is available from the authors upon reasonable request.

For the proposed detectors we fix all hyperparameters before evaluation and keep them constant across all experimental scenarios. The number of training rounds is set to $n=5$, a decision justified by the ablation study in Section \ref{sec:ablation} (Figures \ref{fig:ablation_cfpt} and \ref{fig:ablation_tabautodrift}). The \textit{CFPT} pseudo-labeling confidence threshold is set to $\tau = 0.9$, as it is commonly used in semi-supervised pseudo-labeling according to \cite{9}, \cite{10} and \cite{11}. The drift decision threshold is set at $\theta = 0.5$ for both detectors, as this is the natural midpoint of the utility score ($U \in [0,1]$). We emphasize that $\theta$ is not an internal tuning constant but a reliability/cost operating point. A lower $\theta$ favours reliability by flagging drift more readily, whereas a higher $\theta$ favours cost by signalling retraining more conservatively. Because the utility score is bounded and label‐free, this operating point can be set and adjusted online by the MLOps pipeline, according to the retraining budget and accuracy target of a given deployment, without post‐deployment ground truth. This behaviour contrasts with the window sizes, forgetting factors, and sensitivity thresholds of classical detectors, which are not calibrated to a reliability/cost axis and must be re‐tuned for each deployment.

\subsection{Performance metrics}
\label{perf_metric}
In the following subsections, we use the $D_1$ - $D_{30}$ stemming from the \textit{fingerprinting} and the $D_1$ - $D_8$ stemming from the \textit{"links"} datasets to evaluate our detectors and to assess whether they are able to identify the concept drift phenomenon accurately. Since the detectors proposed in this work use a utility function that designates whether the $M_0$ model should be retrained or not, it is difficult to compare them with existing solutions. This happens because each method uses a different prediction confidence threshold that is uniquely tailored for itself. For this reason, we define the following reward function that can be commonly used among several techniques, to compare their efficiency:
\begin{equation}
\label{eq:performance-metric}
    Reward=\begin{cases}
    F1_{gain},      & \text{if decision is true positive}.\\
    T_s,      & \text{if decision is true negative}.\\
    F1_{gain}-T_s,  & \text{if decision is false positive}.\\
    -F1_{gain},  & \text{if decision is false negative}.\\
    \end{cases}
\end{equation}
The reward favours correct decisions and penalizes incorrect ones, scaled by the drift magnitude. A correct alarm is rewarded by the macro-F1 gain $F1_{gain}$ that retraining would yield; a correct non-alarm by a small user-set constant $T_s$ (set to $0.1$ in our experiments); a false alarm is penalized by $F1_{gain}-T_s$, which is negative when no real drift is present and $F1_{gain}$ is therefore small; and a missed drift by $-F1_{gain}$.

We should notice that in order to find the value of the $F1_{gain}$, during the experimental evaluation, we generate the ground-truth labels $L_1^{gt}$ for the $D_1$ dataset manually. This way we can assess how much will the F1 score of the model $M_0$ be increased, if re-trained with the $D_1$. Such labels are required only for the evaluation procedure and not for the decision-making process of either \textit{TabAutoDrift} or \textit{CFPT}, which operate without requiring the ground-truth labels of $D_1$. In our experiments, we apply the formula described above for each method under examination and for each dataset. At the end of each experiment, we sum each method's rewards and we calculate the final score.

Additionally, for each method, we record whether it raises an "alarm" indicating the need for model retraining. Since our detectors work on batches, our ground truth are datasets in which we know concept drift occurred. By comparing each method's alarms against these ground truths, we can classify outcomes as true positives (correct alarm), false positives (incorrect alarm), true negatives (correct non-alarm), and false negatives (missed alarm). From these counts, we compute precision, recall, and F1-score for each detection method, providing a robust evaluation of its ability to identify drifts.

\subsection{Comparison with existing techniques}
\label{baseline_description}
We compare the techniques proposed in this work with state-of-the art methods analyzed in Section \ref{sec:related}. We perform this comparison with three standardized drift detection method groups, namely \textit{change detection}, \textit{statistical process control} and \textit{window based}. \textit{Change detection} methods leverage statistical tests to assess whether the probability distribution of a stochastic process or time series changes over time. In this paper we utilise the Page-Hinkley \cite{page-cmp} and Cumulative Sum (CUSUM) \cite{page-cmp} techniques from the \textit{change detection} method group. \textit{Statistical process control} techniques are frequently used to control an industrial process or a production method. They use statistical tests which are combined with drift thresholds that determine drifts in a stream of data. We deploy the Drift detection technique (DDM) \cite{ddm-cmp} as a baseline from the \textit{statistical process control} method group. \textit{Window-based} methods analyse data within a pre-defined window, which can be viewed as a temporal frame within a larger dataset or stream. The \textit{window-based} methods we use are the statistical test of equal proportions (STEPD) \cite{stepd-cmp} and the adaptive windowing (ADWIN) \cite{adwin-cmp} techniques. 

In this work, we assume a label-scarce deployment setting and we make sure that each baseline detector operates without requiring any post-deployment labels. Page-Hinkley, CUSUM and ADWIN inherently support this functionality, as they investigate the existence of drift over two different input data streams (the reference $D_0$, with $D_1$, $D_2$, $D_3$ etc.) without consuming any labels. On the other hand, DDM and STEPD require labels to operate since they evaluate the error rate of the $M_0$ model in order to assess the drift levels. To overcome this limitation, we generate pseudo-labels $L_1$ for the dataset under consideration (e.g. $D_1$, $D_2$, $D_3$ etc.) using the $M_0$ model and the technique described in Section~\ref{sec:CFPT}. 
Then, we use the following formula to produce a surrogate signal that represents the raw classification error:  $z_t = 1 - \max P(M_0(D,L_1))$, where $D$ is the input dataset ($D_1$, $D_2$, $D_3$ etc.), $L_1$ the generated pseudo-labels, and $P$ is $M_0$ model's confidence score for the prediction. Evidently, when $P$ drops the error $z_t$ rises which translates to less model confidence for each prediction. The $z_t$ is then fed to the corresponding detector along with the dataset under examination and the reference $D_0$ and $L_0$ to detect drifts as per Eq. \ref{eq:distr_uneq}. Because such information is derived only by the input data (and from the frozen $M_0$'s outputs that generate $L_1$), none of the baselines consumes $L_1^{gt}$ labels at monitoring time; instead they detect drifts in a  label free fashion, exactly like the detectors proposed in this work.


Each baseline requires several user-set parameters, such as window sizes, detection thresholds, and forgetting or delta factors; their per-method symbols and values are reported in Table~\ref{tab:baseline-param}. We implement the detectors under examination using the python library provided by Frouros \cite{CESPEDESSISNIEGA2024101733}. Table \ref{tab:baseline-param} depicts the parameters for each method. For every baseline we adopt the values reported in its original publication and in established comparative studies of drift detectors~\cite{ddm-cmp}, \cite{adwin-cmp}, \cite{stepd-cmp}, \cite{CESPEDESSISNIEGA2024101733}, \cite{sck-ml}, \cite{GONCALVES20148144} and \cite{BARROS2018348}.

\begin{table}[thb]
    \centering
\caption{ The values of the parameters for each method under examination.}
\label{tab:baseline-param}
\resizebox{.49\textwidth}{!}{
  \begin{tabular}{c|c}
  Baseline detectors       & Parameters        \\
    \hline
Page-Hinkley            & $I = 30$, $\delta = 0.005$, $\lambda = 50$, $\alpha = 0.999$    \\
CUSUM                  & $I = 30$, $\delta = 0.005$, $\lambda = 50$   \\ 
DDM                    & $I = 30$, $M = 2$, $D = 3$  \\
STEPD                  & $I = 30$, $W = 0.05$, $D = 0.003$    \\
ADWIN                  & $I = 10$, $W = 5$, $M = 5, \delta=0.002, C= 32$  \\
\hline
CFPT                   & $n = 5$, $\theta = 0.5$, $\tau = 0.9$, XGBoost defaults \cite{12}    \\
TabAutoDrift            & $n = 5$, $\theta = 0.5$, TabNet defaults \cite{13}   \\
\end{tabular}
}
\end{table}

\begin{figure}[t]
    \centering
    \includegraphics[width=0.49\textwidth]{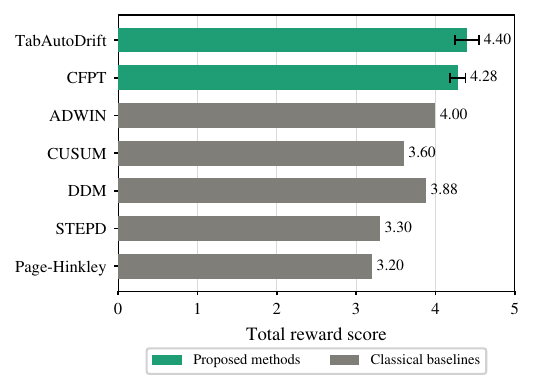}
    \caption {The performance of different concept drift detection detectors using the \textit{fingerprinting} dataset. The \textit{TabAutoDrift} and \textit{CFPT} techniques achieve the best scores, followed by ADWIN and DDM.}
    \label{fig:other-spring}
\end{figure}

\begin{table*}[thb]
    \centering
\caption{Performance of the evaluated detectors on the retraining alarm task, measured by Precision, Recall, and F1-score using the \textit{fingerprinting} dataset.}
\label{tab:spring-pred}
\small
  \begin{tabular}{c|ccccccc}
  detectors               & True positives       & True negatives    & False positives     & False negatives & Precision & Recall & F1 \\
  \hline
  \textbf{TabAutoDrift}              & $9/10$   & $20/20$    & $0/20$    & $1/10$    & $1.00$    & $0.90$ & $0.94$ \\ 
  \textbf{CFPT}                      & $8/10$   & $20/20$    & $0/20$    & $2/10$    & $1.00$    & $0.80$ & $0.88$  \\
 Page-Hinkley              & $6/10$   & $12/20$    & $8/20$    & $4/10$    & $0.42$ & $0.60$ & $0.49$  \\
  CUSUM                     & $7/10$   & $15/20$    & $5/20$    & $3/10$    & $0.58$ & $0.70$ & $0.63$ \\ 
  DDM                       & $7/10$   & $17/20$    & $3/20$    & $3/10$    & $0.70$  & $0.70$ & $0.70$ \\
  STEPD                     & $6/10$   & $13/20$    & $7/20$    & $4/10$    & $0.46$ & $0.60$ & $0.52$ \\
  ADWIN                     & $6/10$   & $16/20$    & $4/20$    & $4/10$    & $0.60$  & $0.60$ & $0.60$\\
\end{tabular}
\end{table*}

\section{Results}
\label{sec:results}

In this section we present the performance results of our detectors compared to the classical detectors as per the methodology in Section~\ref{sec:methodology}. Overall, our detectors match or outperform the classical detectors on both datasets, with CFPT attaining the best results on the \textit{links} dataset and TabAutoDrift the best on the \textit{fingerprinting} dataset, in both cases with substantially fewer false positive retraining triggers.


\subsection{Performance with fingerprinting dataset}
\label{sec:results_finger}

Figure \ref{fig:other-spring} presents a comparison between our \textit{CFPT} and \textit{TabAutoDrift} detectors and with state-of-the-art predictors that exist in the literature. To test our detectors, we run $200$ experiments, using the same configuration options, and we measure the mean, higher, and lower performance. On the other hand, the state-of-the art detectors produce the same results in every experiment, since they are designed to assess whether two input distributions differ and thus, their output is deterministic.

The best performing method is the \textit{TabAutoDrift}, that achieves a total reward of $4.4$. The \textit{CFPT} technique follows with $4.28$, the ADWIN with $4$, the DDM with $3.88$ and the CUSUM with $3.6$. In terms of performance variation, both \textit{TabAutoDrift} and \textit{CFPT} deviate slightly from their mean values by $0.15$ and $0.1$ correspondingly.

Table~\ref{tab:spring-pred} reports, for each detector, the true and false retraining triggers together with the resulting precision, recall, and F1 score. The principal result is that both proposed detectors attain an F1 score of $0.94$ (\textit{TabAutoDrift}) and $0.88$ (\textit{CFPT}) while raising no false retraining alarms, whereas the strongest classical detector, DDM, attains only $0.70$ and the remaining baselines fall between $0.49$ and $0.63$. In a setting in which both missed drifts and unwarranted retraining incur operational cost, the ability to maintain zero false positives while capturing the majority of drifts constitutes a decisive advantage, as it minimizes wasted retraining overhead without sacrificing drift sensitivity.

\begin{figure}[!t]
    \centering
    \includegraphics[width=0.49\textwidth]{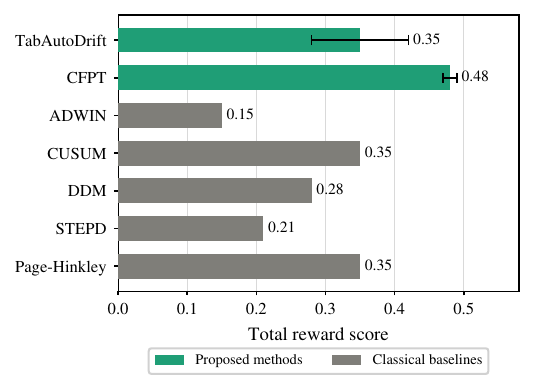}
    \caption {The performance of different concept drift detectors using the \textit{links} dataset. \textit{CFPT} outperforms the rest of the methods by a large margin, followed by TabAutoDrift, Page-Hinkley and CUSUM.}
    \label{fig:other-ts}
\end{figure}

\begin{table*}[!thbp]
    \centering
\caption{Performance of the evaluated detectors on the retraining alarm task, measured by Precision, Recall, and F1-score using the \textit{links} dataset.}
\label{tab:ts-pred}

    \small
  \begin{tabular}{c|ccccccc}
  detectors               & True positives       & True negatives    & False positives     & False negatives & Precision & Recall & F1\\
  \hline
  \textbf{TabAutoDrift}              & $4/5$   & $2/3$    & $1/3$    & $1/5$ &  $0.80$  & $0.80$ & $0.80$  \\ 
  \textbf{CFPT }                     & $5/5$   & $3/3$    & $0/3$    & $0/5$ &  $1.00$    & $1.00$   & $1.00$\\
  Page-Hinkley                & $4/5$   & $2/3$    & $1/3$    & $1/5$ &  $0.80$  & $0.80$ & $0.80$ \\
  CUSUM                     & $4/5$   & $2/3$    & $1/3$    & $1/5$ &  $0.80$  & $0.80$ & $0.80$ \\
  DDM                       & $4/5$   & $1/3$    & $2/3$    & $1/5$ &  $0.66$ & $0.80$ & $0.72$  \\
  STEPD                     & $3/5$   & $1/3$    & $2/3$    & $2/5$ &  $0.60$  & $0.60$ & $0.60$ \\
  ADWIN                     & $2/5$   & $1/3$    & $2/3$    & $3/5$ &  $0.50$  & $0.40$ & $0.44$ \\

  \end{tabular}
\end{table*}

\begin{figure}[t]
    \centering
    \includegraphics[width=0.49\textwidth]{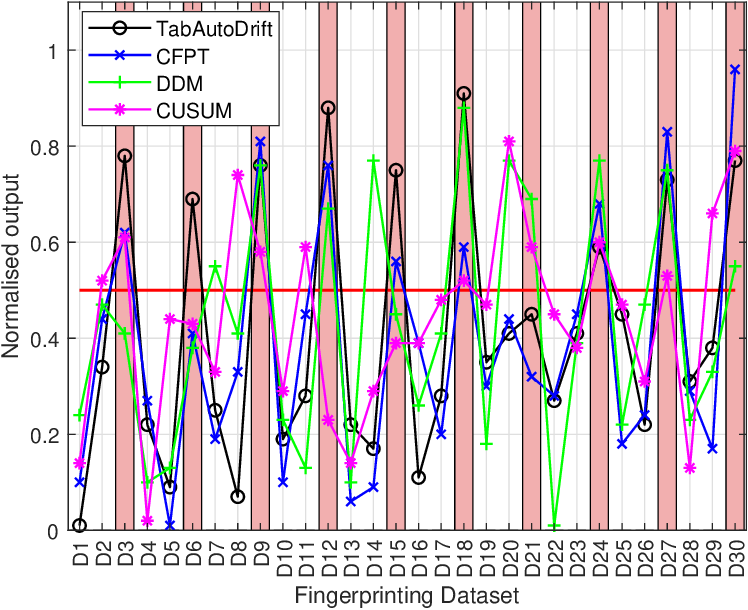}
    \caption {Detailed evaluation of the best performing concept drift detectors for the \textit{fingerprinting} dataset. Even though TabAutoDrift achieves the highest F1-score, there is an instance (in $D_{30}$ dataset) where other methods outperform TabAutoDrift thus, confirming that there is no one size that fits all detector.}
    \label{fig:detail_fing}
\end{figure}
\begin{figure}[t]
    \centering
    \includegraphics[width=0.49\textwidth]{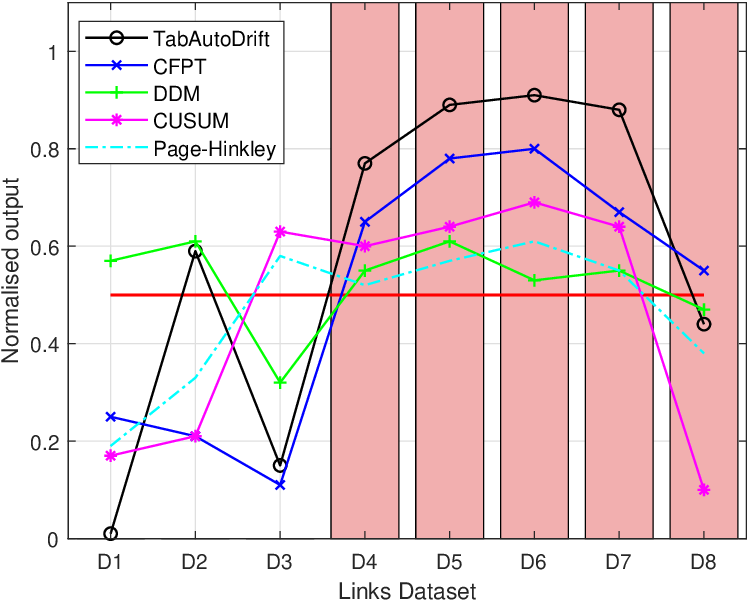}
    \caption {Detailed evaluation of the best performing detectors for the \textit{links} dataset. The proposed \textit{CFPT} technique outperforms the rest of the detectors.}
    \label{fig:detail_ts}
\end{figure}

\subsection{Performance on links dataset}

Figure \ref{fig:other-ts} reports the average reward scores on the \textit{links} dataset along with their corresponding variance. Similarly with the \textit{TabAutoDrift's} evaluation, we run $200$ experiments and we present the mean, best and worst score for our detectors. Results indicate that the detectors proposed in this work substantially outperform classical methods. In particular, the \textit{CFPT} achieves a mean reward of $0.48$ which is $0.13$ points higher than the best traditional detector, while \textit{TabAutoDrift} achieved a mean  reward of $0.35$ and matches Page-Hinkley and CUSUM. The variance for the \textit{CFPT} is $0.01$, whereas for the \textit{TabAutoDrift} is $0.07$. As a result, the \textit{CFPT} technique outperforms the rest of the detectors, even in the worst case scenario. By contrast, the worst case scenario of the \textit{TabAutoDrift} achieves a score of $0.28$, which is lower than the corresponding scores of CUSUM and Page-Hinkley. The DDM scores $0.28$, the STEPD $0.21$ and the ADWIN $0.15$. These results demonstrate that the \textit{CFPT} method triggers alarms at a better time than the SotA methods, ensuring high model performance when concept drift occurs.

Table \ref{tab:ts-pred} summarizes detection accuracy on the \textit{links} dataset. \textit{CFPT} achieves perfect discrimination, with no false alarms and no missed drifts and an F1 score of $1.00$, \textit{TabAutoDrift} matches the strongest classical methods, namely Page-Hinkley and CUSUM, at an F1 score of $0.80$, and the remaining baselines trail between $0.44$ (ADWIN) and $0.72$ (DDM). These results reinforce that the \textit{CFPT} approach delivers superior sensitivity and specificity relative to classical drift detectors.

\subsection{Detailed evaluation under different scenarios}
Figures~\ref{fig:detail_fing} and~\ref{fig:detail_ts} show the per-instance normalized confidence outputs of the best detectors on the $30$ \textit{fingerprinting} and the $8$ \textit{links} datasets. Drifted instances are marked, and the red line denotes the detection threshold ($0.5$), above which a method flags drift; each method's raw confidence is normalized to $[0,1]$ for comparison.

Consistent with Section~\ref{sec:results_finger}, \textit{TabAutoDrift} leads on \textit{fingerprinting} and \textit{CFPT} on \textit{links}. The per-instance view also shows that no single detector dominates every case: \textit{TabAutoDrift} misses the drift in $D_{21}$ that DDM and CUSUM catch, yet is the only method to flag $D_6$; and on \textit{links}, DDM alone correctly rejects the non-drifted $D_3$ that CUSUM and Page-Hinkley misflag.

\subsection{Ablation study for the effects of training and retraining on the expected utility}
\label{sec:ablation}

In this section, we present an ablation study to analyze the sensitivity of our proposed detectors, \textit{CFPT} and \textit{TabAutoDrift}, to their respective training hyperparameters. For the \textit{CFPT} method, we specifically investigate how the number of initial training rounds and subsequent re-training rounds influences the final utility function, while the rest of the hyperparameters are set to default. Concurrently, for the \textit{TabAutoDrift} detector, we evaluate the effect of varying its number of training epochs, while the architecture remain as proposed by the original authors. The primary goal of this analysis is to understand the performance trade-offs associated with these parameters and to offer empirical guidance for their optimal configuration. The ablation study was conducted on a \textit{fingerprinting} dataset, since as per Section~\ref{sec:methodology}-A it has significantly higher number of samples compared to the \textit{links} dataset.

\begin{figure}[!thbp]
    \centering
    \includegraphics[width=0.49\textwidth]{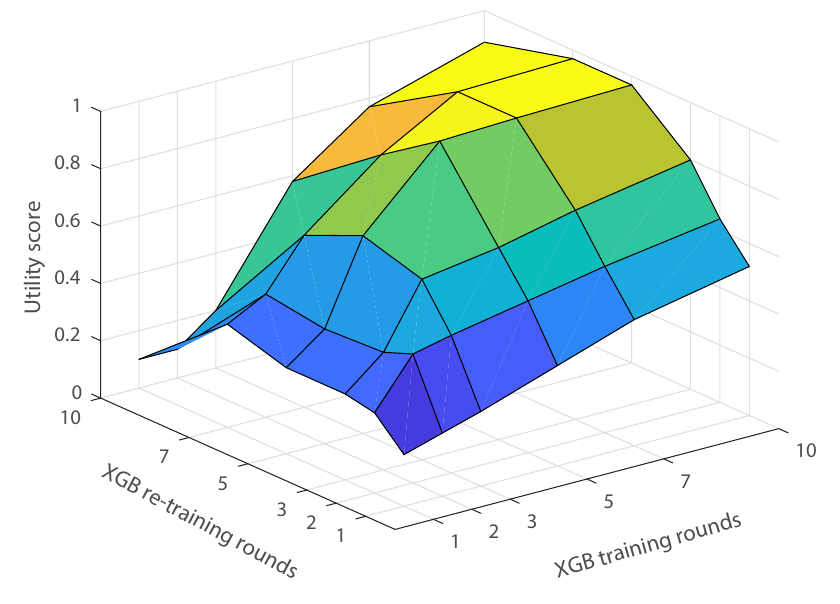}
    \caption {The effect of different number of training and re-training rounds to the \textit{CFPT} method's utility score.}
    \label{fig:ablation_cfpt}
\end{figure}

In Figure~\ref{fig:ablation_cfpt} the effects of different number of training and re-training rounds to the \textit{CFPT's} utility score can be seen. For the purposes of this illustration, we have chosen two datasets where the concept drift phenomenon is prominent, to properly evaluate our findings. Thus, we expect the method to produce high utility scores i.e., clear assessments that the original model needs to be trained with the new dataset. Results indicate that more training rounds result in higher utility scores. This is expected since using larger training periods, the XGBoost learns the $D_0$ representations better and converges faster to a decision regarding the existence of concept drift. On the other hand, choosing a training period of over $5$ rounds marginally improves the utility score; instead it increases the computational complexity of the \textit{CFPT}. The results that consider the XGBoost re-training rounds  tell a different story. There, an optimal parameter space exists, outside of which, the \textit{CFPT's} utility score drops by a large margin. Since the XGBoost re-training operation tries to learn the representations of the new data (e.g. $D_1$, $D_2$, $D_3$ etc.), a large number of re-training rounds risks of properly training the XGBoost model with the new dataset. In such cases, the XGBoost model starts fitting the new data and then, a lower utility score is produced since no deviations are observed from the newly arrived data. As a result, the optimal number of re-training rounds is empirically set to $5$ in order to prevent the XGBoost model to fit the new data.

\begin{figure}[t]
    \centering
    \includegraphics[width=0.35\textwidth]{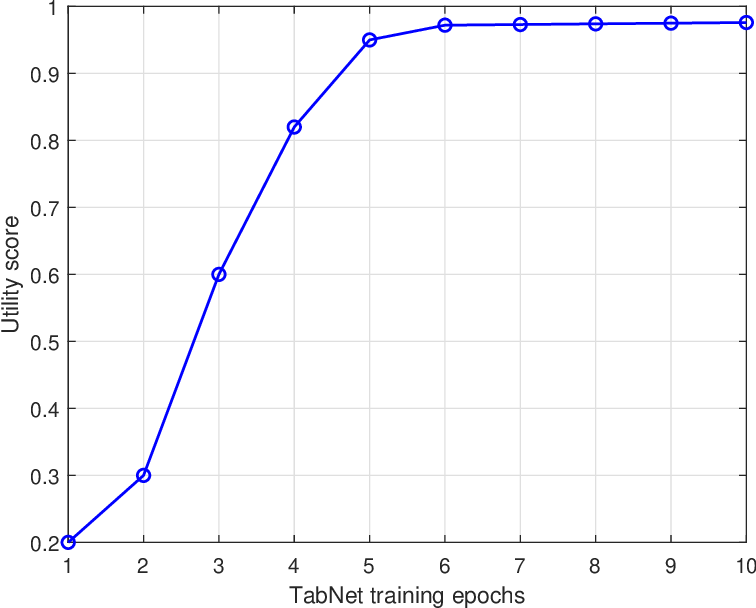}
    \caption {The effect of different number of training epochs to the \textit{TabAutoDrift}  method's utility score.}
    \label{fig:ablation_tabautodrift}
\end{figure}

We perform a similar study for the impact of different number of training epochs to the \textit{TabAutoDrift} method's utility score and we showcase its results in Figure~\ref{fig:ablation_tabautodrift}. Again, we select two datasets in which the concept drift phenomenon exists, and we expect a high utility score to properly detect this. Evidently, a comparable phenomenon appears for TabNet as well. The increase of training epochs increases the expected utility but up to a certain point. After $5$ training epochs the increase is marginal and does not justify the imposed computational overhead.

\subsection{Execution time requirements}
In Table \ref{tab:exec-time} we present the execution time requirements of each method, in seconds. We also report the standard deviation of each method's execution time. Generally, the methods that exist in the literature are very fast, since they complete their assessments within $5$ seconds for the \textit{fingerprinting} dataset or within $2$ seconds for the \textit{links} dataset. On the other hand, the techniques proposed in this work have larger execution times. \textit{TabAutoDrift} requires $18.5$ (for the \textit{fingerprinting} dataset)  and $15.2$ (for the \textit{links} dataset). \textit{CFPT} completes its execution within $13.3$  and $8.7$ seconds correspondingly. This is expected, as these techniques (partially) train AI models, which severely affects their execution times. Nonetheless, gradual drifts usually occur in longer periods of time, from several hours or several months, as noted by~\cite{costa2025analysis}, so our detectors' higher execution time can still be regarded as relatively fast and affordable for real-world applications.

\begin{table}[!htbp]
    \centering
\caption{Execution times of concept drift detectors, using the \textit{fingerprinting} and \textit{links} datasets.}
\label{tab:exec-time}
\small
  \begin{tabular}{c|cccccc}
  detectors               & \textit{fingerprinting} &  \textit{links}\\
  \hline
  \textbf{TabAutoDrift }           & $18.5s \pm 1.2s$  & $15.2s \pm 1s$         \\ 
  \textbf{CFPT}                    & $13.3s \pm 0.8s$  & $8.7s \pm 0.6s$        \\ 
  Page-Hinkley                       & $4.1s \pm 0.4s$  & $1.4s \pm 0.2s$    \\
  CUSUM                            & $4.1s \pm 0.2s$  & $0.84s \pm 0.02s$         \\ 
  DDM                              & $4s \pm 0.1s$  & $0.1s \pm 0.001s$        \\
  STEPD                            & $4.1s \pm 0.2s$  & $0.03s \pm 0.001s$         \\
  ADWIN                            & $5.4s \pm 0.3s$  & $1.9s \pm 0.1s$        \\
  \end{tabular}

\end{table}

\subsection{Qualitative comparison}
The parameters of \textit{TabAutoDrift} and \textit{CFPT} ($n, \tau, \theta$) are transportable and are not fine-tuned between the two datasets, whereas the classical baselines each expose several window, sensitivity, and forgetting parameters that must be re-tuned per deployment (Table~\ref{tab:baseline-param}). The advantage of our detectors stems from the richer representations learned by TabNet and XGBoost: unlike standard detectors that operate on raw statistical or distance measures, TabNet uses sequential feature selection with sparse masks to learn task-relevant representations, while XGBoost forms a nonlinear ensemble that partitions the input space, so both capture subtler data characteristics and generalize better than traditional methods.

\subsection{Limitations and future work}
Our study has several limitations that also outline directions for future work.
First, both detectors operate on \emph{batches} rather than sample-by-sample
streams, so detection latency is bounded by the batch size; adapting the
expected-utility test to a truly online setting is left for future work.
Second, the detectors partially train an XGBoost or TabNet model at detection
time, incurring higher execution times ($8.7$--$18.5$\,s) than the classical
baselines; although this remains acceptable for the hours-to-months timescale
of gradual wireless drift~\cite{costa2025analysis}, it may be limiting in
latency- or energy-constrained deployments. Third, in both wireless testbeds drift
is induced by the emergence of new classes; validation on additional drift
mechanisms, for instance pure covariate or prior-probability shift, and on
non-wireless benchmarks would further substantiate the generality that the design
targets. Finally, our detectors raise the retraining alarm but do not perform the
retraining itself, nor do they prescribe how much post-drift data to collect before
retraining, as these actions are the responsibility of the surrounding MLOps
pipeline. Closing the loop, by coupling the component to an automated retraining and
data-selection policy, is a natural next step.

\section{Conclusion}
\label{sec:conclusion}

Maintaining the reliability of deployed machine‐learning models under distribution drift is a fundamental requirement for trustworthy AI, and it is especially pressing in non‐stationary, label‐scarce settings such as AI‐native wireless systems, in which drift must be detected without labels so that an MLOps pipeline can retrain and redeploy the affected models. To address this requirement, we cast drift handling as a feedback control loop and presented two complementary, model‐external concept‐drift detectors, \textit{CFPT} and \textit{TabAutoDrift}, that realize the \emph{Observer}: they monitor the deployed model (the System Under Observation) and signal an MLOps \emph{Actuator} when it should be retrained. Both detectors are model‐agnostic and label‐free when deployed. By introducing an expected‐utility metric to trigger retraining, our approach balances sensitivity to true drift against the cost of unnecessary updates. Here, label-free refers to the post-deployment data: both detectors are trained on the originally available labels $L_0$ and require no ground-truth labels for the newly arriving data. Extensive experiments on fingerprinting‐based localization and link‐level anomaly datasets show that our detectors achieve state‐of‐the‐art F1‐scores of up to $0.94$ in localization and up to $1.00$ in anomaly detection, significantly outperforming traditional statistical and window‐based methods. 


\bibliographystyle{IEEEtran}
\bibliography{references.bib}

@ARTICLE{1,
  author={Lu, Jie and Liu, Anjin and Dong, Fan and Gu, Feng and Gama, João and Zhang, Guangquan},
  journal={IEEE Transactions on Knowledge and Data Engineering}, 
  title={Learning under Concept Drift: A Review}, 
  year={2019},
  volume={31},
  number={12},
  pages={2346-2363},
  doi={10.1109/TKDE.2018.2876857}}

@incollection{2,
    author = {Amos, Storkey},
    isbn = {9780262170055},
    title = {23When Training and Test Sets Are Different: Characterizing Learning Transfer},
    booktitle = {Dataset Shift in Machine Learning},
    publisher = {The MIT Press},
    year = {2008},
    month = {12},
    doi = {10.7551/mitpress/9780262170055.003.0001},
    url = {https://doi.org/10.7551/mitpress/9780262170055.003.0001},
    eprint = {https://academic.oup.com/mit-press-scholarship-online/book/0/chapter/166918395/chapter-ag-pdf/44903282/book\_13447\_section\_166918395.ag.pdf},
}

@article{3,
  title={On the Impact of Industrial Delays when Mitigating Distribution Drifts: an Empirical Study on Real-world Financial Systems},
  author={Simonetto, Thibault and Cordy, Maxime and Ghamizi, Salah and Le Traon, Yves and Lefebvre, Cl{\'e}ment and Boystov, Andrey and Goujon, Anne},
  journal={KDD Workshop},
  year={2024}
}

@ARTICLE{4,
  author={Luan, Siyu and Gu, Zonghua and Freidovich, Leonid B. and Jiang, Lili and Zhao, Qingling},
  journal={IEEE Access}, 
  title={Out-of-Distribution Detection for Deep Neural Networks With Isolation Forest and Local Outlier Factor}, 
  year={2021},
  volume={9},
  number={},
  pages={132980-132989},
  doi={10.1109/ACCESS.2021.3108451}}

@article{5,
title = {Deep neural network to extract high-level features and labels in multi-label classification problems},
journal = {Neurocomputing},
volume = {413},
pages = {259-270},
year = {2020},
issn = {0925-2312},
doi = {https://doi.org/10.1016/j.neucom.2020.06.117},
url = {https://www.sciencedirect.com/science/article/pii/S0925231220311115},
author = {Marilyn Bello and Gonzalo Nápoles and Ricardo Sánchez and Rafael Bello and Koen Vanhoof}
}

@InProceedings{7,
author="Wang, Zhixiong
and Wang, Wei",
editor="Liang, Qilian
and Wang, Wei
and Mu, Jiasong
and Liu, Xin
and Na, Zhenyu
and Chen, Bingcai",
title="Concept Drift Detection Based on Kolmogorov--Smirnov Test",
booktitle="Artificial Intelligence in China",
year="2020",
publisher="Springer Singapore",
address="Singapore",
pages="273--280",
isbn="978-981-15-0187-6"
}

@article{8,
  author    = {Jeomoan Francis Kurian and Mohamed Allali},
  title     = {Detecting drifts in data streams using Kullback-Leibler (KL) divergence measure for data engineering applications},
  journal   = {Journal of Data, Information and Management},
  volume    = {6},
  number    = {3},
  pages     = {207--216},
  year      = {2024},
  month     = {September},
  doi       = {10.1007/s42488-024-00119-y},
  url       = {https://doi.org/10.1007/s42488-024-00119-y},
  issn      = {2524-6364}
}

@misc{9,
      title={Why the pseudo label based semi-supervised learning algorithm is effective?}, 
      author={Zeping Min and Qian Ge and Cheng Tai},
      year={2023},
      eprint={2211.10039},
      archivePrefix={arXiv},
      primaryClass={cs.LG},
      url={https://arxiv.org/abs/2211.10039}, 
}

@INPROCEEDINGS{10,
  author={Arazo, Eric and Ortego, Diego and Albert, Paul and O’Connor, Noel E. and McGuinness, Kevin},
  booktitle={2020 International Joint Conference on Neural Networks (IJCNN)}, 
  title={Pseudo-Labeling and Confirmation Bias in Deep Semi-Supervised Learning}, 
  year={2020},
  volume={},
  number={},
  pages={1-8},
  doi={10.1109/IJCNN48605.2020.9207304}}

@misc{11,
      title={In Defense of Pseudo-Labeling: An Uncertainty-Aware Pseudo-label Selection Framework for Semi-Supervised Learning}, 
      author={Mamshad Nayeem Rizve and Kevin Duarte and Yogesh S Rawat and Mubarak Shah},
      year={2021},
      eprint={2101.06329},
      archivePrefix={arXiv},
      primaryClass={cs.LG},
      url={https://arxiv.org/abs/2101.06329}, 
}

@inproceedings{12,
author = {Chen, Tianqi and Guestrin, Carlos},
title = {XGBoost: A Scalable Tree Boosting System},
year = {2016},
isbn = {9781450342322},
publisher = {Association for Computing Machinery},
address = {New York, NY, USA},
url = {https://doi.org/10.1145/2939672.2939785},
doi = {10.1145/2939672.2939785},
booktitle = {Proceedings of the 22nd ACM SIGKDD International Conference on Knowledge Discovery and Data Mining},
pages = {785–794},
numpages = {10},
location = {San Francisco, California, USA},
series = {KDD '16}
}

@misc{13,
      title={TabNet: Attentive Interpretable Tabular Learning}, 
      author={Sercan O. Arik and Tomas Pfister},
      year={2020},
      eprint={1908.07442},
      archivePrefix={arXiv},
      primaryClass={cs.LG},
      url={https://arxiv.org/abs/1908.07442}, 
}

@INPROCEEDINGS{14,
  author={Bertalanič, Blaž and Morano, Grega and Cerar, Gregor},
  booktitle={2022 International Balkan Conference on Communications and Networking (BalkanCom)}, 
  title={LOG-a-TEC Testbed outdoor localization using BLE beacons}, 
  year={2022},
  volume={},
  number={},
  pages={115-119},
  doi={10.1109/BalkanCom55633.2022.9900607}}

@article{15,
  author = {Massey, Frank J.},
  title = {The Kolmogorov-Smirnov Test for Goodness of Fit},
  journal = {Journal of the American Statistical Association},
  volume = {46},
  number = {253},
  year = {1951},
  pages = {68--78},
  doi = {10.2307/2280095},
  url = {https://doi.org/10.2307/2280095},
  note = {Accessed 13 Mar. 2025}
}

@ARTICLE{9715175,
  author={Bertalanič, Blaž and Meža, Marko and Fortuna, Carolina},
  journal={IEEE Transactions on Neural Networks and Learning Systems}, 
  title={Resource-Aware Time Series Imaging Classification for Wireless Link Layer Anomalies}, 
  year={2023},
  volume={34},
  number={10},
  pages={8031-8043},
  doi={10.1109/TNNLS.2022.3149091}}

@article{liu2023leaf,
  title={Leaf: Navigating concept drift in cellular networks},
  author={Liu, Shinan and Bronzino, Francesco and Schmitt, Paul and Bhagoji, Arjun Nitin and Feamster, Nick and Crespo, Hector Garcia and Coyle, Timothy and Ward, Brian},
  journal={Proceedings of the ACM on Networking},
  volume={1},
  number={CoNEXT2},
  pages={1--24},
  year={2023},
  publisher={ACM New York, NY, USA}
}

@article{moreno2012unifying,
  title={A unifying view on dataset shift in classification},
  author={Moreno-Torres, Jose G and Raeder, Troy and Alaiz-Rodr{\'\i}guez, Roc{\'\i}o and Chawla, Nitesh V and Herrera, Francisco},
  journal={Pattern recognition},
  volume={45},
  number={1},
  pages={521--530},
  year={2012},
  publisher={Elsevier}
}

@article{CESPEDESSISNIEGA2024101733,
title = {Frouros: An open-source Python library for drift detection in machine learning systems},
journal = {SoftwareX},
volume = {26},
pages = {101733},
year = {2024},
issn = {2352-7110},
doi = {https://doi.org/10.1016/j.softx.2024.101733},
url = {https://www.sciencedirect.com/science/article/pii/S2352711024001043},
author = {Jaime {Céspedes Sisniega} and Álvaro {López García}}
}

@inproceedings{costa2025analysis,
  title={Analysis of Descriptors of Concept Drift and Their Impacts},
  author={Costa, Albert and Giusti, Rafael and dos Santos, Eulanda M},
  booktitle={Informatics},
  volume={12},
  number={1},
  pages={13},
  year={2025},
  organization={MDPI}
}

@INPROCEEDINGS{saimler10597102,
  author={Saimler, Merve and İckin, Selim and Bernini, Giacomo and Toumi, Nassima and Diamanti, Maria and Papavassiliou, Symeon and Zivkovic, Milan and Akgul, Ozgur Umut and Khorsandi, Bahare M.},
  booktitle={2024 Joint European Conference on Networks and Communications \& 6G Summit (EuCNC/6G Summit)}, 
  title={The Role of AI Enablers in Overcoming Impairments in 6G Networks}, 
  year={2024},
  volume={},
  number={},
  pages={913-918},
  doi={10.1109/EuCNC/6GSummit60053.2024.10597102}}

@ARTICLE{liu10670198,
  author={Liu, Qiong and Zhang, Tianzhu and Hemmatpour, Masoud and Qiu, Han and Zhang, Dong and Chen, Chung Shue and Mellia, Marco and Aghasaryan, Armen},
  journal={IEEE Communications Magazine}, 
  title={Operationalizing AI/ML in Future Networks: A Bird's Eye View from the System Perspective}, 
  year={2025},
  volume={63},
  number={4},
  pages={176-182},
  doi={10.1109/MCOM.001.2400033}}

@ARTICLE{wilhelmi10726906,
  author={Wilhelmi, Francesc and Szott, Szymon and Kosek-Szott, Katarzyna and Bellalta, Boris},
  journal={IEEE Communications Magazine}, 
  title={Machine Learning and Wi-Fi: Unveiling the Path Toward AI/ML-Native IEEE 802.11 Networks}, 
  year={2024},
  volume={},
  number={},
  pages={1-7},
  doi={10.1109/MCOM.001.2400292}}

@article{yungaicela2024misconfiguration,
  title={Misconfiguration in O-RAN: Analysis of the impact of AI/ML},
  author={Yungaicela-Naula, Noe M and Sharma, Vishal and Scott-Hayward, Sandra},
  journal={Computer Networks},
  pages={110455},
  year={2024},
  publisher={Elsevier}
}

@article{page-cmp,
 ISSN = {00063444},
 URL = {http://www.jstor.org/stable/2333009},
 author = {E. S. Page},
 journal = {Biometrika},
 number = {1/2},
 pages = {100--115},
 publisher = {[Oxford University Press, Biometrika Trust]},
 title = {Continuous Inspection Schemes},
 urldate = {2025-06-12},
 volume = {41},
 year = {1954}
}

@InProceedings{ddm-cmp,
author="Gama, Jo{\~a}o
and Medas, Pedro
and Castillo, Gladys
and Rodrigues, Pedro",
editor="Bazzan, Ana L. C.
and Labidi, Sofiane",
title="Learning with Drift Detection",
booktitle="Advances in Artificial Intelligence -- SBIA 2004",
year="2004",
publisher="Springer Berlin Heidelberg",
address="Berlin, Heidelberg",
pages="286--295",
isbn="978-3-540-28645-5"
}

@InProceedings{stepd-cmp,
author="Nishida, Kyosuke
and Yamauchi, Koichiro",
editor="Corruble, Vincent
and Takeda, Masayuki
and Suzuki, Einoshin",
title="Detecting Concept Drift Using Statistical Testing",
booktitle="Discovery Science",
year="2007",
publisher="Springer Berlin Heidelberg",
address="Berlin, Heidelberg",
pages="264--269",
isbn="978-3-540-75488-6"
}

@inbook{adwin-cmp,
author = {Albert Bifet and Ricard Gavaldà},
title = {Learning from Time-Changing Data with Adaptive Windowing},
booktitle = {Proceedings of the 2007 SIAM International Conference on Data Mining (SDM)},
chapter = {},
pages = {443-448},
doi = {10.1137/1.9781611972771.42},
URL = {https://epubs.siam.org/doi/abs/10.1137/1.9781611972771.42},
eprint = {https://epubs.siam.org/doi/pdf/10.1137/1.9781611972771.42}
}

@inproceedings{jaworski2020concept,
  title={Concept drift detection using autoencoders in data streams processing},
  author={Jaworski, Maciej and Rutkowski, Leszek and Angelov, Plamen},
  booktitle={International Conference on Artificial Intelligence and Soft Computing},
  pages={124--133},
  year={2020},
  organization={Springer}
}

@inproceedings{yuan2022recent,
  title={Recent Advances in Concept Drift Adaptation Methods for Deep Learning.},
  author={Yuan, Liheng and Li, Heng and Xia, Beihao and Gao, Cuiying and Liu, Mingyue and Yuan, Wei and You, Xinge},
  booktitle={IJCAI},
  pages={5654--5661},
  year={2022}
}

@article{SHYAA2024109143,
title = {Evolving cybersecurity frontiers: A comprehensive survey on concept drift and feature dynamics aware machine and deep learning in intrusion detection systems},
journal = {Engineering Applications of Artificial Intelligence},
volume = {137},
pages = {109143},
year = {2024},
issn = {0952-1976},
doi = {https://doi.org/10.1016/j.engappai.2024.109143},
url = {https://www.sciencedirect.com/science/article/pii/S0952197624013010},
author = {Methaq A. Shyaa and Noor Farizah Ibrahim and Zurinahni Zainol and Rosni Abdullah and Mohammed Anbar and Laith Alzubaidi},
}

@article{nitish2024class,
  title={Class imbalance and concept drift invariant online botnet threat detection framework for heterogeneous IoT edge},
  author={Nitish, A and Hanumanthappa, J and SP, Shiva Prakash and others},
  journal={Computers \& Security},
  volume={141},
  pages={103820},
  year={2024},
  publisher={Elsevier}
}

@article{chu2024intrusion,
  title={Intrusion detection in the IoT data streams using concept drift localization},
  author={Chu, Renjie and Jin, Peiyuan and Qiao, Hanli and Feng, Quanxi},
  journal={AIMS mathematics},
  volume={9},
  number={1},
  pages={1535--1561},
  year={2024}
}

@inproceedings{andresini2021insomnia,
  title={Insomnia: Towards concept-drift robustness in network intrusion detection},
  author={Andresini, Giuseppina and Pendlebury, Feargus and Pierazzi, Fabio and Loglisci, Corrado and Appice, Annalisa and Cavallaro, Lorenzo},
  booktitle={Proceedings of the 14th ACM workshop on artificial intelligence and security},
  pages={111--122},
  year={2021}
}

@article{suarez2023survey,
  title={A survey on machine learning for recurring concept drifting data streams},
  author={Su{\'a}rez-Cetrulo, Andr{\'e}s L and Quintana, David and Cervantes, Alejandro},
  journal={Expert Systems with Applications},
  volume={213},
  pages={118934},
  year={2023},
  publisher={Elsevier}
}

@article{soltani2024multi,
  title={A multi-agent adaptive deep learning framework for online intrusion detection},
  author={Soltani, Mahdi and Khajavi, Khashayar and Jafari Siavoshani, Mahdi and Jahangir, Amir Hossein},
  journal={Cybersecurity},
  volume={7},
  number={1},
  pages={9},
  year={2024},
  publisher={Springer}
}

@article{GONCALVES20148144,
title = {A comparative study on concept drift detectors},
journal = {Expert Systems with Applications},
volume = {41},
number = {18},
pages = {8144-8156},
year = {2014},
issn = {0957-4174},
doi = {https://doi.org/10.1016/j.eswa.2014.07.019},
url = {https://www.sciencedirect.com/science/article/pii/S0957417414004175},
author = {Paulo M. Gonçalves and Silas G.T. {de Carvalho Santos} and Roberto S.M. Barros and Davi C.L. Vieira}
}

@article{sck-ml,
author = {Montiel, Jacob and Read, Jesse and Bifet, Albert and Abdessalem, Talel},
title = {Scikit-multiflow: a multi-output streaming framework},
year = {2018},
issue_date = {January 2018},
publisher = {JMLR.org},
volume = {19},
number = {1},
issn = {1532-4435},
journal = {J. Mach. Learn. Res.},
month = jan,
pages = {2915–2914},
numpages = {5}
}

@article{BARROS2018348,
title = {A large-scale comparison of concept drift detectors},
journal = {Information Sciences},
volume = {451-452},
pages = {348-370},
year = {2018},
issn = {0020-0255},
doi = {https://doi.org/10.1016/j.ins.2018.04.014},
url = {https://www.sciencedirect.com/science/article/pii/S0020025518302743},
author = {Roberto Souto Maior Barros and Silas Garrido T. Carvalho Santos}
}

@inproceedings{gozuacik2019d3,
  author = {G\"{o}z\"{u}a\c{c}{\i}k, \"{O}mer and B\"{u}y\"{u}k\c{c}ak{\i}r, Alican and Bonab, Hamed and Can, Fazli},
  title = {Unsupervised Concept Drift Detection with a Discriminative Classifier},
  booktitle = {Proc. 28th ACM Int. Conf. on Information and Knowledge Management (CIKM)},
  pages = {2365--2368},
  year = {2019}
}

@inproceedings{rabanser2019failing,
  author = {Rabanser, Stephan and G\"{u}nnemann, Stephan and Lipton, Zachary C.},
  title = {Failing Loudly: An Empirical Study of Methods for Detecting Dataset Shift},
  booktitle = {Advances in Neural Information Processing Systems (NeurIPS)},
  year = {2019}
}

@inproceedings{dosreis2016fast,
  author = {dos Reis, Denis Moreira and Flach, Peter and Matwin, Stan and Batista, Gustavo},
  title = {Fast Unsupervised Online Drift Detection Using Incremental Kolmogorov-Smirnov Test},
  booktitle = {Proc. 22nd ACM SIGKDD Int. Conf. on Knowledge Discovery and Data Mining (KDD)},
  pages = {1545--1554},
  year = {2016}
}

@inproceedings{lipton2018detecting,
  author = {Lipton, Zachary C. and Wang, Yu-Xiang and Smola, Alexander},
  title = {Detecting and Correcting for Label Shift with Black Box Predictors},
  booktitle = {Proc. 35th Int. Conf. on Machine Learning (ICML)},
  pages = {3122--3130},
  year = {2018}
}

@article{gheibi2021applying,
  author  = {Gheibi, Omid and Weyns, Danny and Quin, Federico},
  title   = {Applying Machine Learning in Self-Adaptive Systems: A Systematic Literature Review},
  journal = {ACM Transactions on Autonomous and Adaptive Systems (TAAS)},
  volume  = {15},
  number  = {3},
  articleno = {9},
  year    = {2021}
}

@article{kreuzberger2023mlops,
  author  = {Kreuzberger, Dominik and K\"{u}hl, Niklas and Hirschl, Sebastian},
  title   = {Machine Learning Operations ({MLOps}): Overview, Definition, and Architecture},
  journal = {IEEE Access},
  volume  = {11},
  pages   = {31866--31879},
  year    = {2023}
}

\end{document}